\documentclass[%
twocolumn,
showpacs,preprintnumbers,
amsmath,amssymb,
aps, pra,
]{revtex4-1}
\def\la{{\langle}}
\def\ra{{\rangle}}

\newcommand{\beq}{\begin{equation}}
\newcommand{\eeq}{\end{equation}}
\newcommand{\beqa}{\begin{eqnarray}}
\newcommand{\eeqa}{\end{eqnarray}}

\usepackage{amsmath}
\usepackage{amssymb}
\usepackage{graphicx}
\usepackage{dcolumn}
\usepackage{bm}
\usepackage{color} 
\usepackage{CJK}
\begin{document}
\title{Shortcut to adiabatic population transfer in quantum three-level systems: effective two-level problems and feasible counter-diabatic driving}

\author{Yi-Chao Li}
\affiliation{Department of Physics, Shanghai University, 200444 Shanghai, People's Republic of China}

\author{Xi Chen}
\email[Corresponding author: ]{xchen@shu.edu.cn}
\affiliation{Department of Physics, Shanghai University, 200444 Shanghai, People's Republic of China}

\begin{abstract}
Shortcut to adiabaticity in various quantum systems has attracted much attention with the wide applications in quantum information processing and quantum control. In this paper, we concentrate on stimulated Raman shortcut-to-adiabatic passage in quantum three-level systems. To implement counter-diabatic driving but without additional coupling, we first reduce the quantum three-level systems to effective two-level problems at large intermediate-level detuning, or on resonance, apply counter-diabatic driving along with the unitary transformation, and eventually modify the pump and Stokes pulses for achieving fast and high-fidelity population transfer. The required laser intensity and stability against parameter variation are further discussed, to demonstrate the advantage of shortcuts to adiabaticity.
\end{abstract}


\maketitle
\section{INTRODUCTION}

Coherent manipulation of internal state in various quantum systems plays a significant role in atomic
and molecular physics with the applications in metrology, interferometry, quantum computing, quantum information processing and control of chemical interaction, see review \cite{Bergmann-Rev,Guerin,Shapiro-Rev,Molmer,Bergmann-Rev2}.
Quite often, one of the most important goals is to achieve state preparation or transfer with high fidelity.
So adiabatic approaches such as rapid adiabatic passage (RAP), stimulated Raman adiabatic passage (STIRAP) and their variants \cite{Shapiro-Rev,Molmer,Bergmann-Rev2} have
been proposed and widely applied in different quantum two-level or three-level systems. These adiabatic protocols are robust against the fluctuations of control parameters, as compared to the resonant pulses. However, the long time required to satisfy the adiabatic criteria might be problematic, when the decoherence effect is considered. Therefore, in order to
achieve fast and high-fidelity quantum state control, the optimal control theory \cite{Tannor,Boscain,Kontz,Vitanov-opt} and composite pulses \cite{Levitt,Torosov-PRL,Torosov} have been proposed, by reducing the time consumption and diabatic loss or minimizing systematic errors.

Alternatively,  several works on ``shortcut to adiabaticity" (STA), including counter-diabatic, inverse engineering, and fast-forward approaches,
have been recently devoted to mimicking adiabatic population transfers but within a short time \cite{Berry,Rice,PRL105,ChenPRA11,Ruschhaupt,GuerinPRL,ChenPRA12,Arimondo,Longhi-nonHermite,Sofia,Masuda,SongXK,Xiayan,Clerk}. Among them, the counter-diabatic driving \cite{Rice} (equivalent to transitionless quantum algorithm \cite{Berry}) provides a powerful method
to design complementary interaction appropriately, so that diabatic transition can be suppressed and the system evolves exactly following the adiabatic reference.
Such shortcut protocol has been experimentally demonstrated in (effective) two-level quantum systems, e.g. accelerated optical lattice \cite{Oliver} and
spin of a single NV center in diamond \cite{Suter}. In the three-level atomic systems, additional coupling between initial and target levels can be implemented by a magnetic dipole transition
\cite{PRL105,Unanyan}, which might be problematic in practice. In other systems, it might be even unfeasible.
The way out is to apply the unitary transformation for cancelling the additional coupling but keeping the same dynamics \cite{PRL109,Takahashi,SongXK},
or the (generalized) inverse engineer for pulse shapes \cite{ChenPRA12,Clerk}. However, in many cases  the examples of three-level system on one-photon resonance have been worked out for simplicity.

\begin{figure}[]
\scalebox{0.9}[0.9]{\includegraphics{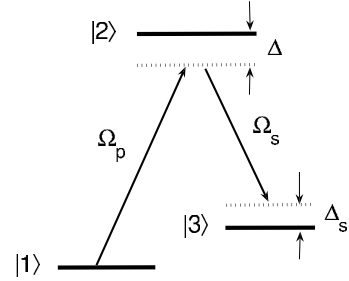}}
\caption{$\Lambda$-type three-level system for STIRAP, where the Rabi frequencies $\Omega_{p,s}$ present the pump
and Stokes pulses, $\Delta$ and $\Delta_s$ are the detunings.}
\label{model}
\end{figure}

In this paper, we shall concentrate on the construction and implementation of stimulated Raman shortcut-to-adiabatic passage (STIRSAP) in quantum three-level systems,
see Fig. \ref{model}. Particularly, large intermediate-level detuning or one-photon resonance are assumed here, since STIRAP in these cases can be reduced to
effective two-level problems \cite{VitanovPRA}. This allows us to utilize the counter-diabatic technique along with unitary transformation
proposed in two-level systems \cite{PRL109}, thus implementing STIRSAP without additional coupling by only modifying pump and Stokes pulses.
In detail, in the case of large detuning, we first reduce the quantum three-level system to an effective two-level system by using ``adiabatic elimination'',
apply counter-diabatic driving along with the unitary transformation, and finally design pump and Stokes pulses. For the sake of completeness,
the counter-diabatic driving in one-photon resonance case and the connection with other shortcut methods is studied. At last,
the stability with respect to the parameter fluctuation is also discussed, showing the advantage of STA. STIRSAP proposed here
can be demonstrated in recent experiments for speeding up SITRAP with cold atom \cite{NC} and solid-state spin systems \cite{NV-3level}.


\section{MODEL and HAMILTONIAN}

Considering the Hamiltonian for STIRAP system within the rotating wave approximation (RWA) \cite{Bergmann-Rev,Bergmann-Rev2,ChenPRA12}
\beq
\label{first}
H_0=\frac{\hbar}{2}
\left(
\begin{array}{ccc}
0 & \Omega_p(t) & 0
\\
\Omega_p(t)&2\Delta&\Omega_s(t)
\\
0&\Omega_s(t)&2\delta_s
\end{array}
\right).
\eeq
Here $\Omega_p(t)$ and $\Omega_s(t)$ are Rabi frequencies of pump and Stokes laser fields, shown in Fig. \ref{model},
where $\Delta=(E_2-E_1)/\hbar-\omega_p$, $\Delta_s=(E_2-E_3)/\hbar-\omega_s$, and $\delta_s=\Delta-\Delta_s$, $\omega_p$ and $\omega_s$ are the laser frequencies of pump and Stokes laser, respectively, and $E_j,j=1,2,3$ are bare-basis state energies. On two-photon resonance ($\delta_s=0$), the Hamiltonian (\ref{first}) reads
\beqa
\label{H0}
H_0=\frac{\hbar}{2}
\left(
\begin{array}{ccc}
0&\Omega_p(t)&0\\
\Omega_p(t)&2\Delta&\Omega_s(t)\\
0&\Omega_s(t)&0
\end{array}
\right),
\eeqa
whose instantaneous eigenstates are
\beqa
\label{eigenstates}
\nonumber
&&|n_{0}  \rangle=  \cos \theta |1\rangle  - \sin \theta |3\rangle,~~
\\ \nonumber
&&|n_{+}  \rangle=  \sin \theta \sin \varphi |1\rangle + \cos \varphi |2\rangle + \cos \theta \sin \varphi |3\rangle,%
\\ \nonumber
&&|n_{-} \rangle=   \sin \theta \cos \varphi |1\rangle  - \sin \varphi |2\rangle + \cos \theta \cos \varphi |3\rangle,
\eeqa
with eigenvalues $E_{+} (t)=\hbar \Omega \cot(\varphi/2)$, $E_{0}=0$, and $E_{-} (t) = -\hbar \Omega \tan(\varphi/2)$.
Two mixing angles are defined by $\tan \theta = \Omega_{p} (t)/ \Omega_{s} (t)$ and $\tan (2 \varphi) = \Omega / \Delta(t)$, with
$\Omega =[\Omega^2_{p}(t) + \Omega^2_{s}(t)]^{1/2}$. The wave functions of this three-level system, $\textbf{c}(t)= [c_1(t),c_2(t),c_3(t)]^{T}$, denoted by $|1 \rangle$, $|2\rangle$ and $|3 \rangle$,
is governed by the time-dependent Schr\"{o}dinger equation $i \hbar d \textbf{c}(t)/dt = H_0 \textbf{c} (t)$.
Once these conditions for adiabatic following, $\dot{\theta} \ll \Omega $ (local) and $\Omega t_f \gg 1$ (global), are satisfied \cite{Bergmann-Rev,Bergmann-Rev2}, the solution of above Schr\"{o}dinger equation coincides with the adiabatic approximation, thus the population can be transferred
$|1 \rangle$ to $|3 \rangle$ along the ``dark state'' $|n_0\rangle$, where $t_f$ is the pulse duration or the so-called operation time.

In order to reproduce STIRAP but within a short time, that is, achieve fast population transfer from $|1\rangle \rightarrow |3 \rangle$,
one can apply the counter-diabatic driving \cite{Rice} (equivalent to quantum transitionless algorithm \cite{Berry,PRL105}),
\beq
\label{cd}
H_{cd} =i\hbar\sum |\partial_t n\ra\la n|,
\eeq
to design the supplementary interaction in the form of \cite{PRL105}
\beq
H_{cd} =\frac{\hbar}{2}
\begin{pmatrix}
0&0&i \Omega_a(t)\\
0&0&0\\
-i \Omega_a (t)&0&0
\end{pmatrix},
\eeq
with
$\Omega_a (t)=2 [\dot{\Omega}_p(t)\Omega_s(t)-\dot{\Omega}_s(t)\Omega_p(t)]/[\Omega_p^2(t)+\Omega_s^2(t)]$.
The additional coupling between $|1\ra$ and $|3 \ra$, implemented by microwave dipole transition in atomic system \cite{Unanyan}, can completely suppress the diabatic transition.
However, such coupling might be difficult or even impossible to implement in various systems. For instance,  the phase mismatch
between laser and microwave fields causes the infidelity \cite{NC}.  Moreover, generating the grating for such coupling is doable in optical multi-mode waveguide,
but not in coupled waveguides \cite{ShuoOE}. In general, when $\Delta \neq 0$ under the two-photon resonance condition,
the cancellation of counter-diabatic interaction becomes more challenging by using unitary transformation, as compared to the case of one-photon resonance ($\Delta=0$),
since the eight Gell-Mann matrices are involved in the dynamics of such three-level systems satisfying SU(3) Lie algebra \cite{Hioe}.
In what follows we shall propose the method  of implementation of counter-diabatic driving in three-level systems.
Following \cite{VitanovPRA}, we reduce STIRAP to the effective two-level problems by considering the adiabatic elimination under large detuning ($\Delta \gg \Omega$)
or one-photon resonance ($\Delta =0$), which enables to implement STIRSAP without additional coupling by using the similar strategy originally proposed
in two-level systems \cite{PRL109}.

As we shall deal with two different examples with the assumption of large detuning and one-photon resonance. It is unavoidable to repeat some
symbols, such as $H_{cd}$, $\Omega_a(t)$, $\theta$, $H_{\textrm{eff}}$, $\tilde{H}_{\textrm{eff}}$, $\Omega_{\textrm{eff}}(t)$, $\Delta_{\textrm{eff}}(t)$,
$\tilde{\Omega}_{\textrm{eff}}(t)$, $\tilde{\Delta}_{\textrm{eff}}(t)$, and $\tilde{\Omega}_{p,s}(t)$.
So consistency is strictly guaranteed only within each case. The detail of how this comes about will be clarified in the context.

\section{Feasible shortcut design}

\subsection{large detuning ($\Delta \gg \Omega$)}
At large intermediate-level detuning, $\Delta \gg \Omega$,
level $|2\rangle$ is scarcely populated ($\dot{c}_2 (t) \simeq 0$), and it can be adiabatically eliminated to obtain the following effective two-level Hamiltonian in the subspace of levels $|1 \rangle$ and $|3\rangle$ \cite{VitanovPRA}:
\beq
\label{Heff1}
H_{\textrm{eff}}= \frac{\hbar}{2}
\begin{pmatrix}
-\Delta_{\textrm{eff}}(t) & \Omega_{\textrm{eff}}(t) \\
\Omega_{\textrm{eff}}(t) &  \Delta_{\textrm{eff}}(t)
\end{pmatrix},
\eeq
where the effective detuning $\Delta_{\textrm{eff}} (t)$ and Rabi frequency $\Omega_{\textrm{eff}}(t)$  are
\beqa
\label{eff1}
\Delta_{\textrm{eff}}(t)=\frac{\Omega_p^2(t)-\Omega_s^2(t)}{4\Delta},\\
\label{eff2}
\Omega_{\textrm{eff}}(t)=-\frac{\Omega_p(t)\Omega_s(t)}{2\Delta}.
\eeqa
\begin{figure}[]
\scalebox{0.7}[0.7]{\includegraphics{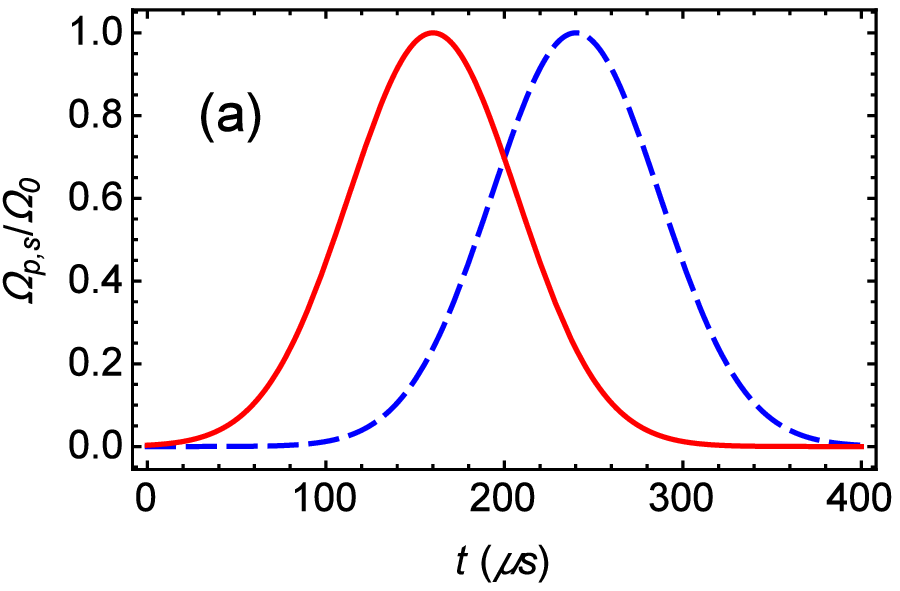}}
\scalebox{0.7}[0.7]{\includegraphics{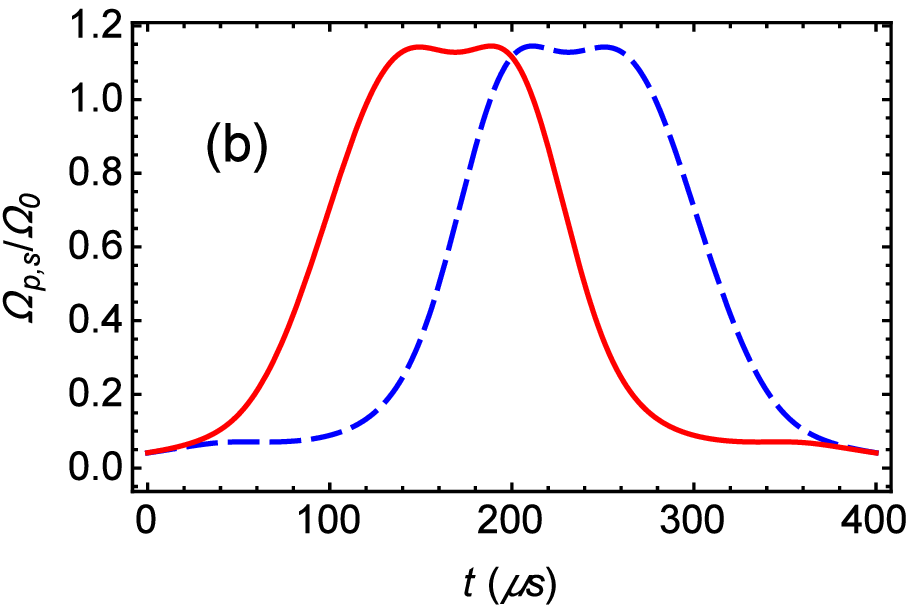}}
\scalebox{0.7}[0.7]{\includegraphics{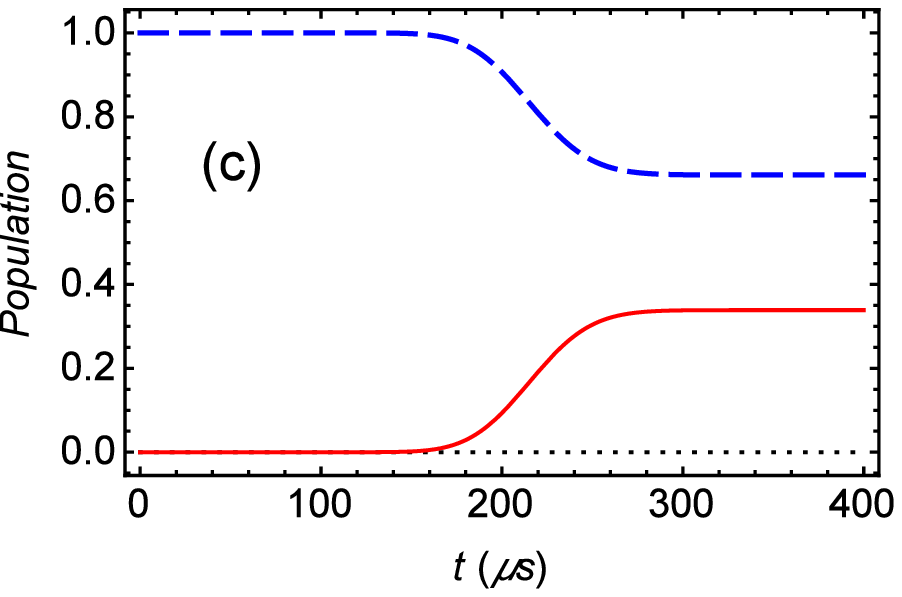}}
\scalebox{0.7}[0.7]{\includegraphics{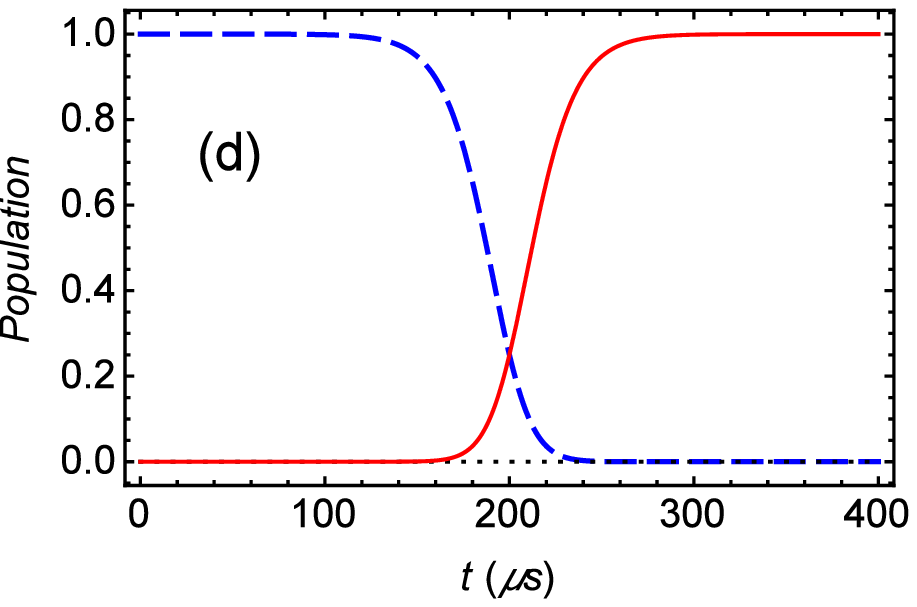}}
\caption{(Color online) Different Rabi frequencies for STIRAP (a) and STIRSAP (b), where
Stokes (solid red) and pump (dashed blue) pulses are shown. The state evolutions of STIRAP (c) and STIRSAP (d) are also compared,
where population of levels $|1\rangle$ (dashed  blue), $|2\rangle$ (dotted black), and $|3\rangle$ (solid red) is presented.
Parameters: $\Omega_0= 2 \pi \times 5$ MHz, $\Delta= 2 \pi \times 2.5$ GHz, $t_f=400$ $\mu$s, $\tau=t_f/10$ and $\sigma=t_f/6$. }
\label{fig2}
\end{figure}

Once the effective two-level Hamiltonian (\ref{Heff1}) is obtained, we can
calculate the counter-diabatic driving, from the definition (\ref{cd}), as \cite{PRL105}
\beq
\label{H1}
H_{cd} =\frac{\hbar}{2}
\begin{pmatrix}
0&-i\Omega_a(t)\\
i\Omega_a(t)&0
\end{pmatrix},
\eeq
where $\Omega_a(t)=[\Omega_\textrm{eff}(t)\dot{\Delta}_\textrm{eff}(t)-\dot{\Omega}_\textrm{eff}(t)\Delta_\textrm{eff}(t)]/[\Delta_\textrm{eff}^2(t)+\Omega_\textrm{eff}^2(t)]$. Assisted by the counter-diabatic term (\ref{H1}), the system can be driven along the adiabatic path of reference Hamiltonian (\ref{Heff1}) within a short time. The total Hamiltonian, $H=H_{\textrm{eff}}+H_{cd}$, is constructed as
\beq
\label{HI}
H=\frac{\hbar}{2}
\begin{pmatrix}
-\Delta_\textrm{eff}(t)&\sqrt{\Omega_\textrm{eff}^2(t)+\Omega_a^{2}(t)}e^{-i\phi}\\
\sqrt{\Omega_\textrm{eff}^2(t)+\Omega_a^{2}(t)}e^{i\phi}&\Delta_\textrm{eff}(t)
\end{pmatrix},
\eeq
where $\phi(t)=\arctan[\Omega_a(t)/\Omega_\textrm{eff}(t)]$. By applying the unitary transformation,
\beq
\label{unitary}
U(t)=
\begin{pmatrix}
e^{-i\phi(t)/2}&0\\
0&e^{i\phi(t)/2}
\end{pmatrix},
\eeq
we can further obtain $\tilde{H}_\textrm{eff}=U^{\dag} H U-i\hbar U^{\dag}\dot{U}$, namely,
\beq
\label{Htotal}
\tilde{H}_\textrm{eff}=\frac{\hbar}{2}
\begin{pmatrix}
- \tilde{\Delta}_{\textrm{eff}}(t)  & \tilde{\Omega}_{\textrm{eff}}(t)  \\
\tilde{\Omega}_{\textrm{eff}}(t) & \tilde{\Delta}_{\textrm{eff}}(t)
\end{pmatrix},
\eeq
with
$
\tilde{\Delta}_{\textrm{eff}}(t) = \Delta_{\textrm{eff}}(t)+\dot{\phi}
$ and
$
\tilde{\Omega}_{\textrm{eff}}(t) =\sqrt{\Omega_{\textrm{eff}}^2(t)+\Omega_a^{2}(t)}
$.
Obviously, the unitary transformation means the rotation along $z$-axis, which results in the cancellation of $\sigma_y$ term in the Hamiltonian (\ref{HI}). In principle,
the population dynamics of Hamiltonian (\ref{Htotal}) is the same as the previous one (\ref{HI}), up to the global phase. When the boundary condition $U(0)=U(t_f)=1$ is satisfied,
the initial and final population is the same as the adiabatic reference. Now let us go back to the three-level system and design the modified pump and Stokes fields
by comparing the Hamiltonian (\ref{Htotal}) and (\ref{Heff1}). Like Eqs. (\ref{eff1}) and (\ref{eff2}), we impose
\beqa
\tilde{\Delta}_{\textrm{eff}}(t) &=& \frac{\tilde{\Omega}_p^2-\tilde{\Omega}_s^2}{4\widetilde{\Delta}},
\\
\tilde{\Omega}_{\textrm{eff}}(t) &=& -\frac{\tilde{\Omega}_p(t) \tilde{\Omega}_s(t)}{2\widetilde{\Delta}},
\eeqa
and calculate inversely the modified fields as
\beqa
\label{tildep}
\tilde{\Omega}_p(t)&=&\sqrt{2\widetilde{\Delta}(\sqrt{\tilde{\Delta}^2_{\textrm{eff}}(t)+\tilde{\Omega}^2_{\textrm{eff}}(t)}+\tilde{\Delta}_{\textrm{eff}}(t))},
\\
\label{tildes}
\tilde{\Omega}_s(t)&=&\sqrt{2\widetilde{\Delta}(\sqrt{\tilde{\Delta}^2_{\textrm{eff}}(t)+\tilde{\Omega}^2_{\textrm{eff}}(t)}-\tilde{\Delta}_{\textrm{eff}}(t))}.
\eeqa
In order to guarantee that the problem of a two-level system with counter-diabatic term can be transformed back to a three-level system with modified Stokes and pumping pulses,
we should have $\widetilde{\Delta} \gg \tilde{\Omega}_{p,s}(t)$. Here it is reasonable to assume $\widetilde{\Delta}=\Delta$, since the original detuning $\Delta$ is the order of GHz, but the (modified) Rabi frequency is the order of MHz, see the parameters in Fig. \ref{fig2}.
Substituting the expressions of $\tilde{\Omega}_{\textrm{eff}}(t)$ and $\tilde{\Delta}_{\textrm{eff}}(t)$ into Eqs. (\ref{tildep}) and (\ref{tildes}), we finally obtain new designed laser fields to drive the state following the dynamics of effective two-level Hamiltonian (\ref{Htotal}), thus implementing STIRSAP at large intermediate-level detuning.

\begin{figure}[]
\scalebox{0.75}[0.75]{\includegraphics{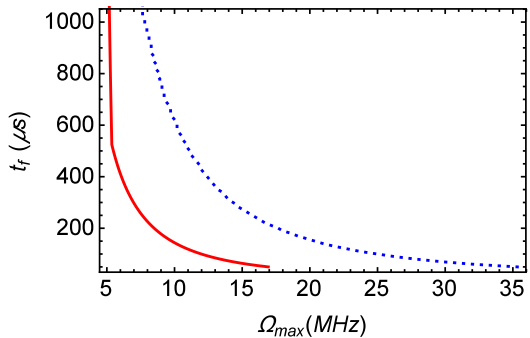}}
\caption{(Color online) Operation time $t_f$ versus the peak value of Rabi frequency $\Omega_{\textrm{max}}$  for STIRSAP (solid red) and STIRAP (dashed blue) when the fidelity is above $99\%$. Parameters
are the same as those in Fig. \ref{fig2}.}
\label{fig3}
\end{figure}

To illustrate how shortcut protocol works for STIRAP, we assume the original pump and Stokes pulses, see Fig. \ref{fig2} (a), as an adiabatic reference,
\beqa
\label{omegap}
\Omega_p(t)=\Omega_0\,\exp\,\left[-\frac{(t-t_f/2-\tau)^2}{\sigma^2}\right],
\\
\label{omegas}
\Omega_s(t)=\Omega_0\,\exp\,\left[-\frac{(t-t_f/2+\tau)^2}{\sigma^2}\right],
\eeqa
with full width at half maximum (FWHM) $\sigma$, separation time between the two pulses $\tau$, and the amplitude $\Omega_0$.
We set the detuning $\Delta=2 \pi \times 2.5$ GHz to guarantee large detuning, $\Delta \gg \Omega_0$, for the validity of ``adiabatic elimination".
In this case, the operation time required for adiabatic process should be larger than resonant $\pi$ pulse, $t_f \gg t_{\pi}= 2\pi \Delta/ \Omega^2_0$.
Under the parameters:  $\Omega_0=2 \pi \times 5$ MHz,  $t_f=400$ $\mu$s ($t_{\pi}=100$ $\mu$s), $\tau=t_f/10$ and $\sigma=t_f/6$, the dynamics of original Hamiltonian
(\ref{H0}) with pump and Stokes pulses, see Eqs. (\ref{omegap}) and (\ref{omegas}), is not adiabatic at all, and population is not completely transferred
from $|1\rangle$ to $|3\rangle$, see Fig. \ref{fig2} (c). By using modified pulses predicted from Eqs. (\ref{tildep}) and (\ref{tildes}), the perfect population transfer can be achieved within a short time,
as shown in Figs. \ref{fig2} (b) and (d).
The shapes of modified pump and Stokes pulse are smooth enough to generate in the experiment with cold atoms \cite{NC}, though they are slightly different from original Gaussian ones.

Importantly, we have to point out the energy cost for the acceleration of STIRAP. By comparing the pulses in STIRAP and STIRSAP, we see that the peak value of modified Rabi frequencies is larger than
original ones. It is reasonable that more laser intensity is required to shorten the operation time, and the relation between energy (laser intensity) and operation time, in general,
satisfies $\Omega_{\textrm{max}} \propto 1/t_f$. To quantify it, we plot the operation time versus maximum value of laser intensity $\Omega_{\textrm{max}}$ in Fig. \ref{fig3}.
Given the fidelity above $99.99\%$, the maximum laser intensity required for shortcuts is less than that for STIRAP. In other words,
when an allowed laser intensity is fixed, the operation time for STIRSAP is less than that for conventional STIRAP. For example,
if the maximum value of laser intensity is $\Omega_{\textrm{max}}= 2\pi \times 10$ MHz, the time required for STIRSAP is $t_f = 145$ $\mu$s, which is about 4.3 times faster
than the original STIRAP, $t_f=620$ $\mu$s.

\subsection{one-photon resonance ($\Delta\equiv0$)}

\begin{figure}[]
\scalebox{0.7}[0.7]{\includegraphics{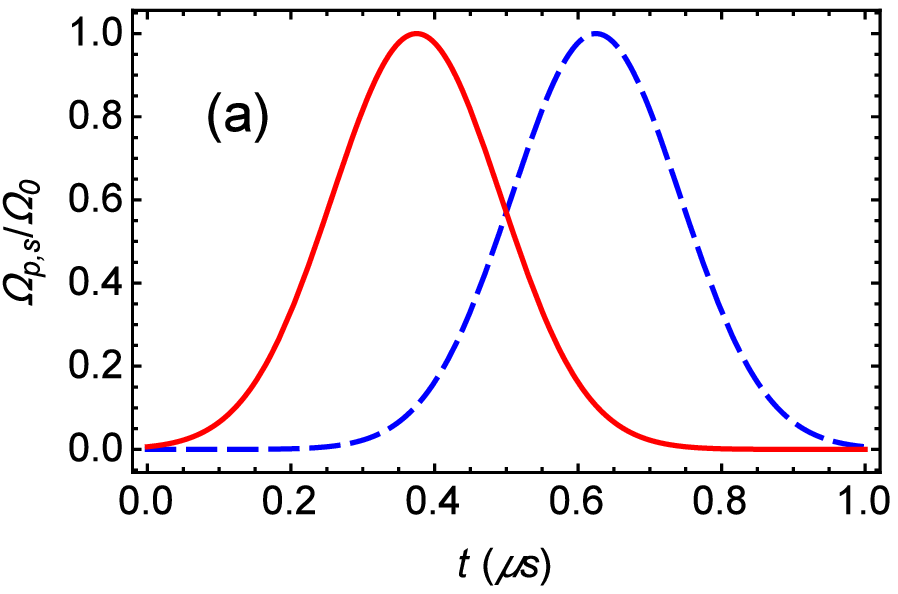}}
\scalebox{0.7}[0.7]{\includegraphics{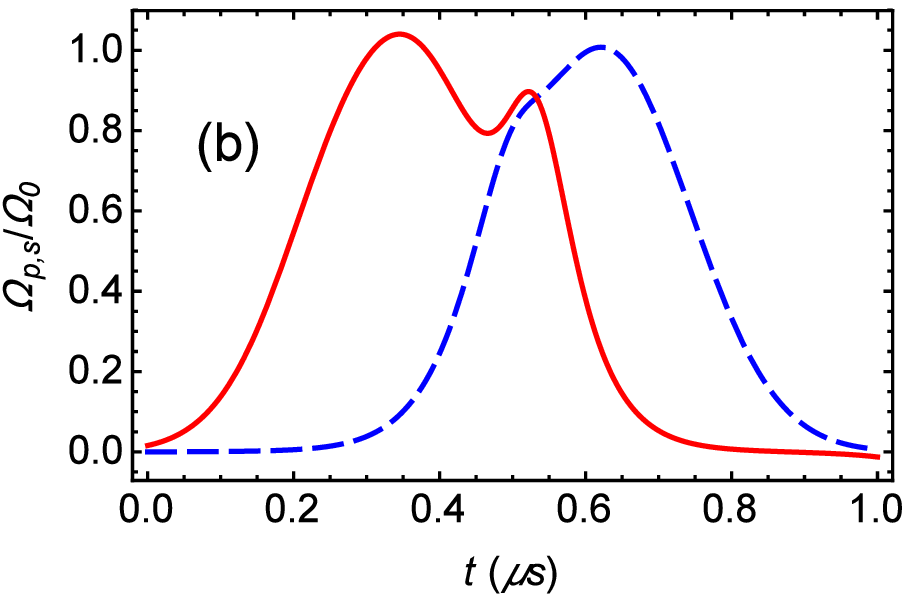}}\\
\scalebox{0.7}[0.7]{\includegraphics{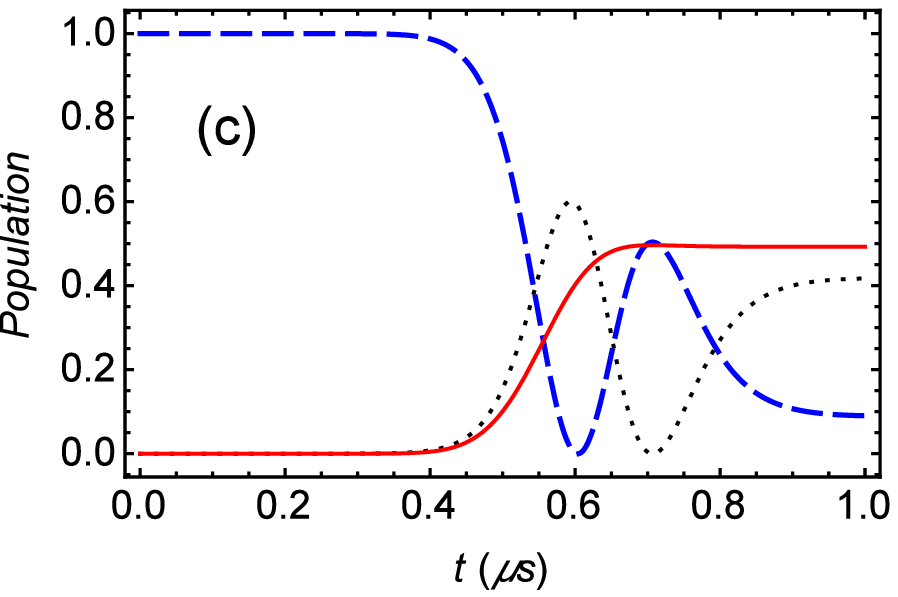}}
\scalebox{0.7}[0.7]{\includegraphics{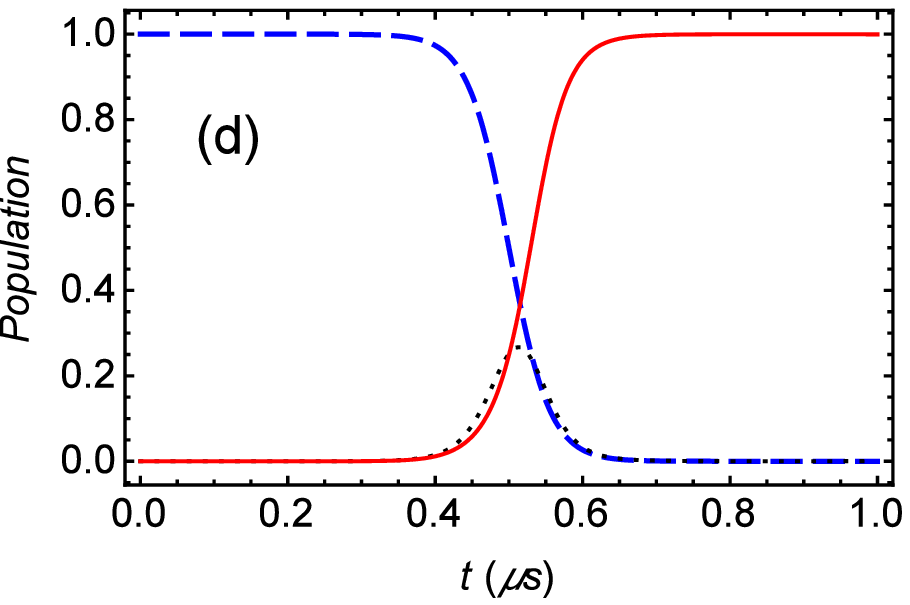}}
\caption{(Color online)  Different Rabi frequencies for STIRAP (a) and STIRSAP (b), where
Stokes (solid red) and pump (dashed blue) pulses are shown. The state evolutions of STIRAP (c) and STIRSAP (d) are also compared,
where population of levels $|1\rangle$ (dashed  blue), $|2\rangle$ (dotted black), and $|3\rangle$ (solid red) is presented.
Parameters: $\Omega_0= 2 \pi \times 5$ MHz, $t_f=1$ $\mu$s, $\tau=t_f/8$ and $\sigma=t_f/6$.}
\label{fig4}
\end{figure}

On one-photon resonance, $\Delta=0$, the three-level system is similarly reduced to the following effective two-level system \cite{VitanovPRA},
\beq
\label{Sch}
H_{\textrm{eff}} =\frac{\hbar}{2}
\begin{pmatrix}
-\Delta_{\textrm{eff}} (t)& \Omega_{\textrm{eff}} (t)\\
\Omega_{\textrm{eff}} (t)& \Delta_{\textrm{eff}} (t)
\end{pmatrix},
\eeq
with effective Rabi frequency and detuing, $\Omega_{\textrm{eff}} (t)= \Omega_p (t)/2$ and $\Delta_{\textrm{eff}} (t)= -\Omega_s (t)/2$.
Noting that the Hamiltonian (\ref{H0}) on resonance ($\Delta =0$) has the same form as
the optical Bloch equations for such effective two-level system (in units $\hbar =1$) \cite{Carroll,Laine}.
The probability amplitudes $c_j(t)$ of the three-level system with Hamiltonian (\ref{H0}) are related to the corresponding two-level amplitudes $b_j(t)$ by
$ c_1(t) = |b_1(t)|^2-|b_2(t)|^2$, $c_2(t) = 2 i \mbox{Im} [b_1^*(t)b_2(t)]$, and $c_3(t) = - 2 \mbox{Re}[b_1^*(t)b_2(t)]$, where $j$ denotes the number of states.
Again, the total Hamiltonian is $H= H_\textrm{eff}+ H_{cd}$, where the counter-diabatic term is calculated as \cite{PRL105}
\beq
H_{cd} =\frac{\hbar}{2}
\begin{pmatrix}
0&-i\Omega_a(t)\\
i\Omega_a(t)&0
\end{pmatrix}.
\eeq
with $\Omega_a(t)=[\dot{\Omega}_p(t)\Omega_s(t)-\dot{\Omega}_s(t)\Omega_p(t)]/[\Omega_p^2(t)+\Omega_s^2(t)]$.
After $z$-axis rotation by using unitary transformation (\ref{unitary}), we can obtain
\beq
\tilde{H}_\textrm{eff}=\frac{\hbar}{2}
\begin{pmatrix}
- \tilde{\Delta}_{\textrm{eff}}(t) & \tilde{\Omega}_{\textrm{eff}}(t)  \\
\tilde{\Omega}_{\textrm{eff}}(t)  & \tilde{\Delta}_{\textrm{eff}}(t)
\end{pmatrix},
\eeq
with the new definition, $\tilde{\Delta}_{\textrm{eff}}(t) = \Delta_{\textrm{eff}}(t)+\dot{\phi}$,
$\tilde{\Omega}_{\textrm{eff}}(t) =[\Omega_{\textrm{eff}}^2(t)+\Omega_a^2(t)]^{1/2}$, and $\phi(t)=\arctan[2\Omega_a(t)/\Omega_p(t)]$.
Supposing the two-level problem can be transformed back to three-level problem, we
can impose $\tilde{\Omega}_{\textrm{eff}}(t) = \tilde{\Omega}_p(t)/2$ and $\tilde{\Delta}_{\textrm{eff}}(t)= -\tilde{\Omega}_s(t)/2$,
and the modified pump and Stokes Rabi frequency can be inversely calculated as
\beqa
\label{ModiRabi-1}
\tilde{\Omega}_p(t)&=& \sqrt{\Omega_p^2(t)+4\Omega_a^2(t)},
\\
\label{ModiRabi-2}
\tilde{\Omega}_s(t)&=& \Omega_s(t)-2\dot{\phi}(t).
\eeqa
Figs. \ref{fig4} (a) and (b) show the new designed pump and Stokes pulses, as compared to the original ones. The evolution
of state in Figs. \ref{fig4} (c) and (d) demonstrates that by using STIRSAP the population transfer can be achieved with fidelity $1$,
while the previous STIRAP does not work perfectly. The parameters are $\Omega_0= 2 \pi \times 5$ MHz, $t_f=1$ $\mu$s, $\tau=t_f/8$ and $\sigma=t_f/6$.
The operation time used here is very short, and not much larger than $t_{\pi}= \sqrt{2}\pi/ \Omega_0 \simeq 0.14$ $\mu$s for resonant $\pi$ pulse.
So the influence of spontaneous emission might be negligible, though the level $|2\rangle$ is populated.

\begin{figure}[]
\scalebox{0.75}[0.75]{\includegraphics{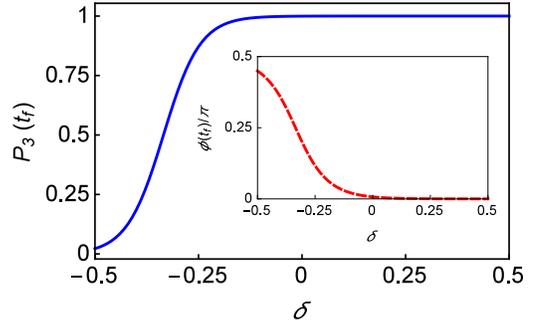}}
\caption{(Color online) Fidelity (solid blue) versus the variation of separation time $(1+\delta)\tau$. Inset: Dependence of phase $\phi$ (solid red) versus $\delta$ for further explanation.
Parameters are the same of those in Fig. \ref{fig4}.}
\label{fig5}
\end{figure}

Fig. \ref{fig5} shows that the final population transfer is sensitive to the variation of separation time $\tau$, described by $(1+\delta)\tau$. When decreasing $\tau$, the fidelity becomes
worse. However, in the case of large detuning, the fidelity is robust against the fluctuation of severation time $\tau$ \cite{NC}.
As a matter of fact, it is relevant to the mapping between two and three-level problems. When the total Hamiltonian  $H= H_\textrm{eff}+ H_{cd}$ for the effective two-level system is transformed back to
the three-level problem, the Hamiltonian will have the direct coupling $\Omega_a(t)$ between level $|1\rangle$ and $|3\rangle$. But after applying the unitary transformation $U$,
the population dynamics of the Hamiltonian $\tilde{H}_{\textrm{eff}}$ is determined by $\widetilde{\mathbf{b}}(t) = [b'_{1} (t) e^{i\phi/2}, b'_2 (t) e^{-i\phi/2}]^{T}$,
where $b'_{j}(t)$ is the probability amplitudes of two-level systems with Hamiltonian $H = H_{\textrm{eff}}+H_{cd}$.
When going back to three-level system, we can calculate the probability amplitudes of three-level problem from
$\widetilde{c}_1(t)=|b'_{1}(t)|^2-|b'_2(t)|^2$, $\widetilde{c}_2(t) = 2 i \mbox{Im} [{b'}^*_{1}(t)b'_{2}(t) e^{-i\phi}]$, and $\widetilde{c}_3(t)= - 2 \mbox{Re}[{b'}^*_{1}(t)b'_{2}(t)e^{-i\phi}]$.
This provides the population of level $|3\rangle$,
\beq
P_3 (t)\equiv |\widetilde{c}_3(t)|^2= \cos^2[\phi(t)].
\eeq
Clearly, when $\phi(t_f)=0$, the full population transfer, $P_3 (t_f)=1$, can be achieved.
This suggests the condition that the two-level problem can be transformed back to the three-level problem on resonant case.
For example, the phase $\phi(t_f)$ saturates to null when $\tau$ increasing, see the inset in Fig. \ref{fig5}, and the final population
becomes irrelevant to the shift of separation time. In fact, the condition, $U(0)=U(t_f)=1$, guarantees that the populations at initial and final time are the same before and after the unitary transformation \cite{PRL109}.
So one can further choose other functions of adiabatic reference, satisfying $\phi(0)=\phi(t_f)=0$, instead of
Eqs. (\ref{omegap}) and (\ref{omegas}).

Furthermore, as we know, three-level system on one-photon resonance satisfies intrinsic SU(2) Lie algebra, which simplifies the shortcut design.
To clarify it, we rewrite the total Hamiltonian, $H=H_0+H_{cd}$, on resonance, as
\beq
H=\frac{1}{2}[\Omega_p(t)\lambda_1+\Omega_s(t)\lambda_6- 2 \Omega_{a}(t)\lambda_5],
\eeq
where the Gell-Mann matrices,
\beqa
\nonumber
\lambda_1 =
\begin{pmatrix}
0&1&0\\
1&0&0\\
0&0&0
\end{pmatrix},
\lambda_5 =
\begin{pmatrix}
0&0&-i\\
0&0&0\\
i&0&0
\end{pmatrix},
\lambda_6=
\begin{pmatrix}
0&0&0\\
0&0&1\\
0&1&0
\end{pmatrix},
\eeqa
are introduced, and satisfy the commutation relation, $[\lambda_1,\lambda_5]=-i\lambda_6$, $[\lambda_5,\lambda_6]=-i\lambda_1$,
and $[\lambda_6,\lambda_1]=-i\lambda_5$. To get rid of the counter-diabatic term, we introduce the unitary transformation $U(t)=e^{-i\phi(t) \lambda_6}$, that is,
\beq
U(t) =
\begin{pmatrix}
1&0&0\\
0&\cos \phi(t) & -i\sin \phi(t)\\
0&-i\sin \phi(t) &\cos \phi(t)
\end{pmatrix},
\eeq
so that the Hamiltonian, $\tilde{H} =U^{\dag}H U-iU^{\dag}\dot{U}$, becomes
\beq
\tilde{H}=\frac{1}{2}[\tilde{\Omega}_p (t)\lambda_1+\tilde{\Omega}_s (t)\lambda_6- \tilde{\Omega}_{a}(t)\lambda_5],
\eeq
where the Rabi frequencies are
\beqa
\label{new2p}
\tilde{\Omega}_p (t) &=& \Omega_p (t)\cos \phi(t)+ 2 \Omega_{a} (t)\sin \phi(t),
\\
\label{new2s}
\tilde{\Omega}_s (t) &=& \Omega_s(t)- 2\dot{\phi}(t),
\\
\tilde{\Omega}_{a} (t) &=& 2 \Omega_{a}(t)\cos \phi(t)- \Omega_p (t)\sin \phi(t).
\eeqa
Imposing $\tilde{\Omega}_{a} (t) =0$ gives $\phi(t)=\arctan[2\Omega_{a}(t)/\Omega_p(t)]$, which exactly
results in Eqs. (\ref{ModiRabi-1}) and (\ref{ModiRabi-2}). By choosing alternative unitary transformation, $U(t)=e^{- i \phi(t) \lambda_1}$, 
we have the modified pump and Stokes pulses accordingly in the form of
\beqa
\tilde{\Omega}_p(t)&=&\Omega_p (t)-2\dot{\phi} (t),
\\
\tilde{\Omega}_s(t)&=&\sqrt{\Omega_s^2 (t)+4\Omega_{a}^2 (t)}.
\eeqa
Clearly, STIRSAP is achieved with the appropriate boundary condition $U(0)=U(t_f)=1$. This argument is consistent with the condition
that the two-level problem can be transformed back to the three-level problem on resonance.
Otherwise, the population can not be exactly transferred from $|1\rangle$ to $|3\rangle$, see Fig. \ref{fig5}, due to the transformation. It turns out that it is not necessary to reduce the three-level system on one-photon resonance to the effective two-level problem,
since the system has symmetry of SU(2). As a matter of fact, we can further choose a general unitary transformation $U(t)=\exp{\{-i [\alpha(t) \lambda_1 + \beta(t) \lambda_5+\gamma(t) \lambda_6]\}}$, and have more flexibility to design the optimal shortcut, similar to the proposal in the literature \cite{Clerk}. This suggests that shortcut methods are mathematically equivalent \cite{ChenPRA11,ChenPRA12},
though their physical implementation is totally different. Besides, high-order iteration can be applied, in terms of superadiabtic concept \cite{SongXK}.

\begin{figure}[]
\scalebox{0.75}[0.75]{\includegraphics{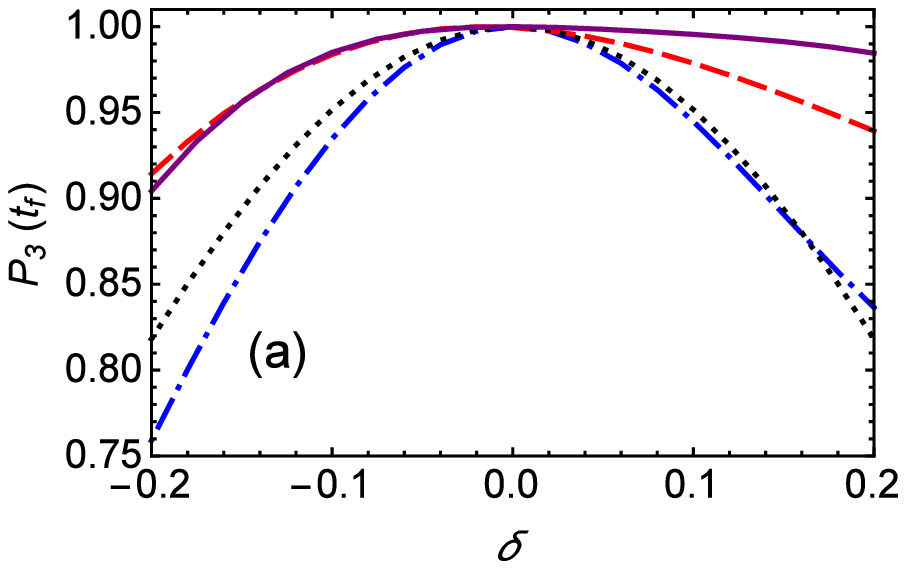}}\\
\scalebox{0.75}[0.75]{\includegraphics{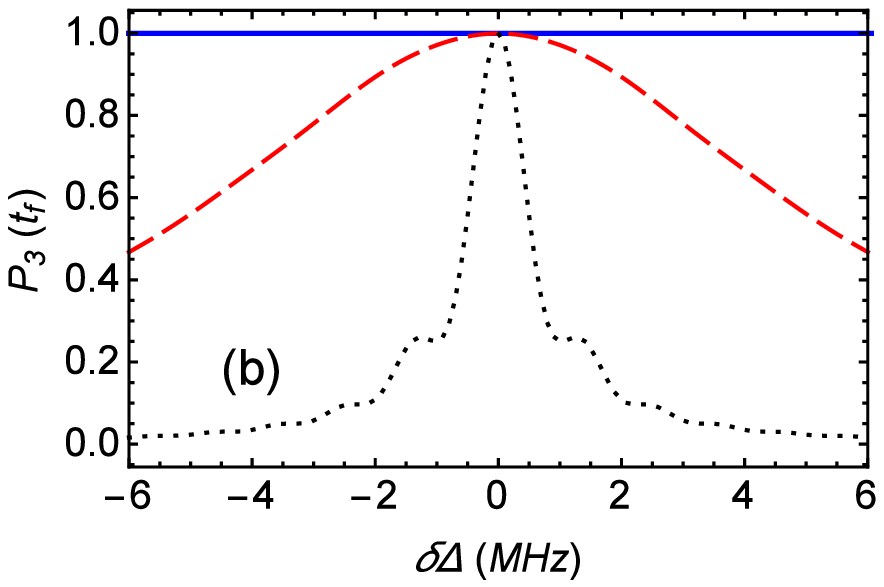}}
\caption{(Color online) (a) Population of level $|3 \rangle$ at final time, $P_3 (t_f)$, versus the variation of laser intensity, where STIRSAP at large detuning: $t_f= 1000$ $\mu$s (solid purple) and $t_f =400$ $\mu$s (dot-dashed blue); STIRSAP on one-photon resonance:  $t_f =1$ $\mu$s (dashed red); resonant $\pi$ pulse: $t_f=1$ $\mu$s (dotted black). (b) $P_3 (t_f)$ versus the variation of detuning, where STIRSAP at large detuning: $t_f= 400$ $\mu$s (solid blue); STIRSAP on one-photon resonance: $t_f =1$ $\mu$s (dashed red);
resonant $\pi$ pulse: $t_f=1$ $\mu$s (dotted black). Other parameters
are the same as those in Figs. \ref{fig2} and \ref{fig4}.}
\label{fig6}
\end{figure}

\section{DISCUSSION}
In this section, we turn to discuss the stability of STIRSAP with respect to different systematic errors. Fig. \ref{fig6} (a) shows that such shortcut protocol on one-photon resonance
is more robust than the resonant $\pi$ pulse (with constant Rabi frequency $\Omega_0= \sqrt{2}\pi/t_f$), when the fluctuation of laser intensity is induced, described by $(1+\delta) \Omega_{s,p}$ or  $(1+\delta) \tilde{\Omega}_{s,p}$. On the contrary, the shortcut protocol at large detuning is not as stable as the shortcut on one-photon resonance and resonant $\pi$ pulse,
especially when operation time is short, $t_f =400$ $\mu$s. However, the stability is improved when $t_f=1000$ $\mu$s and better than other protocols. From the point view of experiment with cold atoms \cite{NC}, STIRSAP at large detuning shows several advantages: (i) level $|2\rangle$ is not populated, which avoids spontaneous decay from excited state; (ii) it is not sensitive to the separation time $\tau$, as compared to the case of one-photon resonance; (iii) resonant $\pi$ pulse does not work perfectly, due to inhomogeneity of atomic cloud \cite{NC}.
Fig. \ref{fig6} (b) also demonstrates STIRSAP is more robust against the detuning error. Particularly, at large detuning the results are not affected by small
vibration of detuning. In addition, we should mention the original STIRAP is accelerated, so the improvement of fidelity by decreasing
operation time is expected, in the presence of spontaneous emission and dephasing noise (calculated by using three-level Lindblad master equation).
Of course, the robustness also depends on the shapes of pump and Stokes fields, and their optimization will be left for further investigation \cite{Ruschhaupt}.

Finally, we shall introduce two kinds of experiments, which are ready to demonstrate our STIRSAP. The parameters through the whole paper are oriented to STIRAP experiment with cold atoms,
where the laser-atom coupling scheme of the three-level are presented, and two ground states $|F=1, m_{F}=0 \rangle = |1 \rangle$ and $|F=2, m_{F}=0 \rangle = |2 \rangle$, and one excited state
$5^2P_{3/2} (=|3\rangle)$ of $^{87}\mbox{Rb}$ are selected as a typical three-level system. One part of the results on STIRSAP at large detuning has been verified in the current experiment \cite{NC}, and definitely
the one-photon resonance case can be tested experimentally as well. On the other hand, the shortcut protocol designed by generalized inverse engineering \cite{Clerk} has been utilized to
control solid-state spin state in NV center \cite{NV-3level}. This $\Lambda$-type three-level system including three spin levels,
$|m_s=0\rangle$, $|m_s=1\rangle$ and $|m_s=-1\rangle$. Such system is available to achieve fast spin manipulation by using STIRSAP with the applications in quantum information processing.

\section{Conclusion}
In summary, we have developed the method to implement STIRSAP without additional coupling by using the counter-diabatic driving in three-level systems.
Considering two cases of large detuning and one-photon resonance, we can reduce the three-level system to an effective two-level problems by using ``adiabatic elimination" or SU(2) Lie algebra. Thereafter, the shapes of pump and Stokes fields are modified to achieve the fast and high-fidelity population transfer without additional coupling between initial and final levels under certain conditions.
This strategy is extremely helpful when we are faced with difficulty in the experiments.
All results can be extended to accelerate the variants of STIRAP, e.g. fractional STIRAP \cite{fSTIRAP}, or
to other adiabatic passages in multi-level systems \cite{Tannor}. The STIRSAP might be also interesting for other analogous quantum three-level systems, see recent review \cite{Mompart-Rev}. 

\section*{Acknowledgment}

This work was supported by NSF of China (Grant No. 11474193), ShuGuang Programm (Grant No. 14SG35), the Specialized Research Fund for the Doctoral Program (Grant No. 2013310811003),
and the Program for Professor of Special Appointment (Eastern Scholar).

\end{document}